%% file: main.tex
\documentclass[journal]{IEEEtran}
\usepackage{hyperref}
\usepackage{amsmath,amssymb,amsthm}
\usepackage{graphics, subfigure}%, amsthm, times, amsfonts}%, cite}
\usepackage{tikz}
\usepackage{pgfplots}
\usepackage{algorithm, algorithmic}
\usepackage[outdir=./]{epstopdf}
\usepackage{epstopdf, xspace}%,cite}
\usepackage{color}
\usepackage{multirow}
\usetikzlibrary{automata,positioning}
\usepackage{wrapfig}
\usepackage{rotating}
\usepackage{flushend}
\usepackage{bbm}

\usepackage{enumitem}
\usepackage{algorithm, algorithmic}

\newtheorem{theorem}{Theorem}

\newcommand{\etc}{\textit{etc.}\xspace}
\newcommand{\ie}{\textit{i.e.,}\xspace}
\newcommand{\eg}{\textit{e.g.,}\xspace}

\usepackage{pgfplots} 
\pgfplotsset{compat=newest} 
\pgfplotsset{plot coordinates/math parser=false} 
\newlength\figureheight 
\newlength\figurewidth 

\usepackage{setspace}
\begin{document}
	\input{macros}

%
% --- Author Metadata here ---
%\conferenceinfo{WOODSTOCK}{'97 El Paso, Texas USA}
%\CopyrightYear{2007} % Allows default copyright year (20XX) to be over-ridden - IF NEED BE.
%\crdata{0-12345-67-8/90/01} % Allows default copyright data (0-89791-88-6/97/05) to be over-ridden - IF NEED BE.
% --- End of Author Metadata ---

\title{FastScan: Robust  Low-Complexity Rate Adaptation Algorithm for Video Streaming over HTTP}

%\author{Anis Elgabli, Vaneet Aggarwal, Ke Liu, and Mark Bell\\email:\{aelgabli, vaneet, liu2067, mrb\}@purdue.edu}

 \author{Anis Elgabli and Vaneet Aggarwal\\Purdue University, West Lafayette, IN}
 	
 	%\thanks{A. Elgabli is with the School of Electrical Computer Engineering,  Purdue University, West Lafayette IN, USA, email: aelgabli@purdue.edu. V. Aggarwal is with the School of Industrial Engineering,  Purdue University, West Lafayette IN, USA, email: vaneet@purdue.edu. }}

\maketitle

\input{abstract}

\begin{IEEEkeywords}Video Streaming, Non-Convex Optimization, Adaptive Bit Rate (ABR),  Average Video Quality, Stall Duration
	\end{IEEEkeywords}

\input{motivation}
%\vspace{-.2in}
%\input{related}
%\vspace{-.0.5in}
\input{problem}

\input{bwPrediction}

\input{algorithm}
\input{implementation}

%\input{simulation_main}
%\input{simulation_main}
%\vspace{-.1in}
\input{conclusion}

\bibliographystyle{IEEEtran} %
%\scriptsize
\bibliography{refs,bib}%,multipath_feng
%\newpage
%\setcounter{page}{1}
%\clearpage
%\appendices

%\input{proofsV3}
\end{document}

%% file: macros.tex
%%%%%%%%%%%%%%%%%%%%%%%%%% COLORS
\definecolor{brown}{cmyk}{0,0.81,1,0.60}
\definecolor{magenta}{rgb}{0.4,0.7,0}
\definecolor{gray}{rgb}{0.5,0.5,0.5}
\definecolor{red}{rgb}{1,0,0}
\definecolor{green}{rgb}{0.5,0,0.5}
\definecolor{blue}{rgb}{0,0,1}

%%%%%%%%%%%%%%%%%%%%%%%%% COMMENTS BY CO-AUTHORS

\ifthenelse{\isundefined{\final}} {
% still in draft mode
%\newcommand{\comment}[1]{{\color{gray}[\textsf{#1}]}}
\newcommand{\vaneet}[1]{{\color{green}(VA: #1)}}
\newcommand{\shuai}[1]{{\color{red}(SH: #1)}}
\newcommand{\feng}[1]{{\color{blue}(FQ: #1)}}
\newcommand{\anis}[1]{{\color{brown}(AE: #1)}}
\newcommand{\shubho}[1]{{\color{magenta}(SS: #1)}}
}{
% in final submission mode
%\newcommand{\comment}[1]{}
\newcommand{\vaneet}[1]{}
\newcommand{\shuai}[1]{}
\newcommand{\feng}[1]{}
\newcommand{\anis}[1]{}
\newcommand{\shubho}[1]{}
}

\newcommand{\eat}[1]{}

\newcommand{\BULLET}{\vspace{+.03in} \noindent $\bullet$ \hspace{+.00in}}

%% file: abstract.tex
%\begin{abstract}
\begin{abstract}
This paper proposes and evaluates a novel algorithm for streaming video over HTTP. The problem is formulated as a  non-convex optimization problem which is constrained by the predicted available bandwidth, chunk deadlines, available video rates, and buffer occupancy. The objective is to optimize a QoE metric that maintains a tradeoff between maximizing the playback rate of every chunk and ensuring fairness among different chunks for the minimum re-buffering time. We propose FastScan, a low complexity algorithm that solves the problem. Online adaptations for dynamic bandwidth environments are proposed with imperfect available bandwidth prediction. Results of  experiments driven by  Variable Bit Rate (VBR) encoded video, video platform system (dash.js), and cellular bandwidth traces of a public dataset reveal the robustness of the online version of FastScan algorithm and demonstrate its significant performance improvement as compared to the considered state-of-the-art video streaming algorithms. For example, on an experiment conducted over 100 real cellular available bandwidth traces of a public dataset that spans different available bandwidth regimes, our proposed algorithm (FastScan) achieves the minimum re-buffering (stall) time and the maximum average playback rate in every single trace as compared to Bola, Festive, BBA, RB, and FastMPC, and Pensieve algorithms.%
	%the state-of-the-art schemes, yet it still achieves the highest average playback rate.% that is $84\%$, $36.2\%$, $36.7$, $17\%$, and $16\%$ higher than the original dash.js rate adaptation scheme, Festive, BBA, RB, and FastMPC, respectively,  with minimum re-buffering (stall) time. 
	%Moreover, DABP does not require a lookup table to be hard coded for videos with different encoding rates as it is a case for FastMPC, yet achieves higher playback rate and less re-buffering time. 
	\end{abstract}

%% file: motivation.tex
%\vspace{-.2in}
\section{Introduction}

%From TV and film to music and sport, streaming has become one of the best ways to get the content you want, when you want it. Instead of buying DVDs, more and more customers prefer to watch TVs and movies online, often on the go on their mobile devices~\cite{atlantic,youtube_stat}. The video streaming industry has even impacted movie theaters: the annual movie tickets bought per person has declined from 4.8 to 4 in the last decade~\cite{theater_survey}. 

The mobile video traffic is estimated to increase by 9x between 2016 and 2021~\cite{cisco_report}.
A convergence of technological, business and social factors are contributing to this  trend. These include the ubiquity of smartphones and tablets, high-speed  cellular connectivity, the increasing availability  of ``over the top'' video content, and a marked  shift in user consumption preferences. Despite numerous adaptive streaming algorithms being devised and deployed, the video quality under mobility is  in many cases unacceptably poor \cite{huawei}. This paper gives a novel algorithm for adaptive video streaming that aims to improve the quality of experience for the end-user. 

In the past decade, a lot of work from both research and industry has focused on the development of  {\em adaptive} video encoding in which the video content on the server side is divided into chunks. Each chunk is then encoded into multiple quality levels and {\em adaptive} video streaming techniques can dynamically adjust the quality of the video being streamed to the  changes in network conditions. The rate adaptive schemes dynamically switch between the different available quality levels based on the network condition,  the client buffer occupancy, \etc 
%The predominant adaptive coding technique in use today is called Adaptive Video Coding~(AVC~\cite{DASH}).
%In AVC, each video chunk is stored into $L$ {\em independent} encoding versions. During playback when fetching a chunk, data corresponding to a certain playback interval in the video,
%the player's adaptation mechanism needs to select one out of the $L$ versions based on its judgment of the network condition and other aforementioned factors.

%Adaptive bit-rate (ABR) video streaming over HTTP and that rely on buffer occupancy, network bandwidth prediction, or both. 

The recent adoption of the open standards MPEG-DASH ~\cite{DASH} has made Adaptive bit-rate (ABR) video streaming the most popular video streaming solution. Commercial systems such as Apples' HLS~\cite{HLS}, Microsoft's Smooth Streaming~\cite{SS}, and Adobe's HDS~\cite{HDS} are all some variants of ABR streaming techniques. In recent studies, researchers have investigated various approaches for making streaming decisions, for example, by using control theory~\cite{MPC,Miller15}, Markov Decision Process~\cite{Jarnikov11}, machine learning~\cite{Claeys14}, client buffer information~\cite{BBA}, and data-driven techniques~\cite{C3,CS2P}. In this paper, we use the prediction of future network condition to provide a novel algorithm for ABR video streaming. 

%The prediction of the future network conditions can play an important role in Internet video streaming. In this paper, we use the prediction of 

%A prior study~\cite{Zou15} investigated the performance gap between the state-of-the-art streaming approaches and the approach with accurate available bandwidth prediction for ABR. The results indicate that prediction brings additional performance boost for ABR, and thus motivates our study.

In \cite{MPC}, a similar optimization problem is considered. However, the proposed optimization formulation in~\cite{MPC} is not shown to be optimally solvable in polynomial complexity. Moreover, a lookup table based on approximating the solution of the offline problem for a given set of encoding rates is proposed. To reduce the table size, the authors suggest dividing the offline solution to bins, so that the final decision is an approximation to the offline solution. However, the table is only valid for a specific set of encoding rates, hence another table need to be generated and stored for a different set of rates. Recently, the authors of \cite{mao2017neural} proposed a system that generates ABR algorithms using reinforcement learning (RL). Their idea is to train a neural network (NN) model that selects bitrates for future video chunks based on observations collected by client's video players. The RL based system uses the metric described in~\cite{MPC} as the QoE metric that the RL agent needs to learn how to optimize. However, such a system needs a lot of training data including videos encoded at different rates and bandwidth traces that span wide range of network bandwidth conditions. %Moreover, their model  uses the throughput that is achieved when downloading last $k$ chunks as an input to the NN model which implicitly provides better predicted bandwidth than harmon.} 

In this paper, we propose a QoE metric that accounts for the diminishing QoE gain as we go higher in the playback rate (quality level), \ie we consider the concave nature of the user's QoE with respect to the playback rate. We formulate the quality decisions of the video chunks as an optimization problem. Moreover, we develop a low complexity algorithm to solve the proposed problem, and we show that the algorithm solves the optimization problem optimally when every quality level is constant Bit Rate (CBR) encoded (\ie when the $n$-th quality level rate is same for all chunks). %Therefore, if a perfect prediction is available for the entire duration of the video, the algorithm provides a theoretic upper bound.
We further show that the algorithm significantly outperform the-state-of-the-art algorithms ABR streaming algorithms. The formulation and the algorithm we propose in this paper are modifications of that proposed formulation and algorithm for SVC (Scalable Video Coding) encoded videos in \cite{AnisTONSingle}.  In this work, we consider AVC encoded videos in which every chunk is encoded into independent versions that represent different qualities. Moreover, in contrast to the work in~\cite{AnisTONSingle}, we implement the proposed algorithm in this paper in a real system with real videos that are AVC encoded.

%the prior ABR streaming algorithms.
 
Since the available bandwidth can only be predicted for short time ahead with prediction error, the proposed algorithm uses the short prediction to make quality decisions for $W$ chunks ahead, and  repeats after the download of every chunk to adjust for prediction errors and to include one more chunk ahead every time. Thus, the algorithm is a sliding window based algorithm. The complexity of the algorithm is linear in $W$, and in contrast to~\cite{MPC}, the approach does not require to pre-store information about different encoding rates.  The main contributions are summarized as follows.

%\subsection{Our Contributions}
%\BULLET
%The main contributions of the paper are as follows.
%\begin{itemize}
 \BULLET We formulate the video streaming over HTTP constrained by the predicted available bandwidth, chunk deadlines, available video rates, and buffer occupancy as a non-convex optimization problem. The objective is to optimize a QoE metric that maintains a tradeoff between maximizing the playback rate of every chunk and ensuring fairness among all chunks for the minimum stall duration.
 
  %whose objective is to  minimize the re-buffering duration as the first priority, and maximize the average quality and minimize the quality switching rate as the second priority.

\BULLET We develop FastScan, a novel low-complexity algorithm to solve the video streaming problem. FastScan has a complexity that is linear in the prediction window size. The algorithm is optimal if the video is CBR encoded. The online adaptation of the algorithm is re-run after the download of every chunk to re-consider and decide the quality of the next window of chunks.

%\BULLET We show that the solution is optimal if the video is CBR encoded. Hence, the offline algorithm, \ie the algorithm that uses perfect available bandwidth prediction for the whole period of the video, provides a genie-aided  bound for the online optimization.

\BULLET Real implementation test bed with the open source video framework dash.js shows significant performance improvement of our algorithm as compared to the-state-of-art algorithms. For example, on an experiment conducted over three sets of real cellular bandwidth traces of  public datasets that spans different available bandwidth regimes, our proposed algorithm (FastScan) achieves the highest QoE in more than $99\%$ of traces in almost every dataset as compared to Festive~\cite{jiang2012improving}, BBA~\cite{BBA}, RB~\cite{MPC}, and Bola~\cite{spiteri2016bola}, FastMPC~\cite{MPC}, and Pensieve~\cite{mao2017neural}. %Moreover, FastScan does not require a lookup table to be hard coded for videos with different encoding rates as it is a case for FastMPC, yet achieves higher playback rate and less re-buffering time. 

The rest of the paper is organized as follows. Section II formulates the problem, and describes the notations. Section III describes the proposed algorithm. Section IV describes the implementation and shows that the proposed algorithm outperforms the state-of-the-art algorithms. Section V concludes the paper. 

%% file: problem.tex
\section{Problem Formulation}
In this section, we describe our ABR streaming problem formulation.  Let's assume a video divided into $V$ chunks (segments), where every chunk is of length $L$ seconds is encoded at one of $(N+1)$ quality levels. Let the $n$-th quality level of a chunk $i$ be encoded at  rate $r_{n,i}$. Let $X_{n,i}$ be the size of the $i$-th chunk when it is encoded at $n$-th quality level, $X_{n,i}=L*r_{n,i}$.  %, where $X_{n,i}\in {\mathcal X}_i = \{Lr_{0,i}, \cdots, Lr_{N,i}\}$. 

%\begin{thisnote}
Let $Y_{n,i}$ be the size difference between the $n$ and $(n-1)$-th quality levels, so $Y_{n,i}=X_{n,i}-X_{n-1,i}$, $n\ge 1$. However, $Y_{0,i}=X_{0,i}$. Moreover, let $I_{n,i}$ be an indicator variable that is equal to $1$ if the $i$-th chunk can be fetched at the $n$-th quality level, and $0$ otherwise, so $I_{n,i}$ is defined as follows:
 \begin{equation}\label{equ:inrVariable}
\left\{\begin{array}{l}
I_{n,i}=1, \text{ if $i$-th chunk can be fetched at $n$-th quality}\\
I_{n,i}=0,  \text{ otherwise}\\
\end{array}\right.
\end{equation}

We will refer to $I_{n,i}$ by the decision variable of the $n$-th quality level. Let $Z_{n,i}$ be equal to $I_{n,i}\cdot Y_{n,i}$, \ie $Z_{n,i}$ is equal to how much is fetched out of $Y_{n,i}$. Since  $I_{n,i} \in \{0,1\}$, $Z_{n,i}$ is  $\in \{0,Y_{n,i}\}$. Since no chunk is totally skipped, we have  $I_{0,i} =1$, and $Z_{0,i}=Y_{0,i} \forall i$.
%\end{thisnote}

We assume an initial start-up delay of $S$ time-units, and the video starts playing at this point. Since each chunk is of duration $L$, chunk $i$ must be downloaded by time $S+(i-1)L$.  If a chunk $i$ cannot be downloaded by $S+(i-1)L$, a stall (\ie rebuffering) of $\alpha(i)$ seconds will occur, and the new deadline of every chunk $i^\prime \geq i$ will increase by $\alpha(i)$ seconds. Where $\alpha(i)$ is the more seconds required to fully download chunk $i$. As stalls continue occur during the video playback, the final deadline of every chunk $i$ will be be $S+(i-1)L+\sum_{k=1}^i \alpha(k)$.\ie $deadline(i)=S+(i-1)L+\sum_{k=1}^i\alpha(k)$. During the stall, the video will pause until the chunk is fully downloaded since the buffer is running empty.

 %, where $r_{n,i}$ is $\in$ ${\mathcal R}_i=\{0,r_{0},r_1,....r_{N}\}$. The size of a chunk that is fetched at $n$th quality level is $X_{n,i}$, where $X_{n,i}=L*r_{n,i}$, and $r_{n,i}$ is the rate of the chunk $i$ when it is fetched at $n$th quality level. 
 %We define $Y_{i,n}$ as the size of the difference between the $(n-1)$th and $n$th quality levels of chunk $i$. i.e, it is the more bits that are needed to take a chunk from $(n-1)$ to $n$th quality size, so $Y_{n,i}=X_{n,i}-X_{n-1,i}$. Moreover, we define our decision variables $Z_{n,i}$ for every quality level $n$ and chunk $i$ as follow:
%\begin{align}
%&Z_{n,i}=Y_{n,i}, \quad \text{if chunk $i$ can be fetched at $n$th quality level}.\nonumber \\
%&0, \quad otherwise
%\end{align}
%$Z_{n,i}$ $\in {\mathcal Z}_n= \{0,Y_{n,i}\}$.
%We will call $Z_{n,i}$ the decision variable of chunk $i$ and $n$th quality level $n$ throughout the paper.
%The final size of a chunk is $X(i)=\sum_{n=0}^N Z_{n,i}$. 

Let $X(i)=\sum_{n=0}^{N} Z_{n,i}$ be the decided size of the $i$-th chunk. %, $X(i) \in {\mathcal X}_i$. In other words, if the $i$-th chunk is decided to be fetched at $n$-th quality level, then $X(i)=X_{n,i}=L*r_{n,i}$. 
Further, let $x(i,j)$ be how much out of $X(i)$ can be fetched at time $j$ and $z_{n}(i,j)$ is what can be fetched out of $Z_{n,i}$ at time $j$. %and let $x(i,j)$ be what can be fetched for chunk $i$ at time slot $j$, i.e. $x(i,j)=\sum_{n=0}^N z_n(i,j)$. 
Moreover, let $B(j)$ be the predicted available bandwidth at time $j$. Also let $B_m$ be the playback buffer size in time units, which indicates the maximum amount of video content that can be held in the playout buffer. We assume, without loss of generality, that the time unit is 1 second. When chunk $i$ starts downloading, the buffer occupancy increases by $L$ seconds (Recall that the chunk size is $L$ seconds).  %The total stall (re-buffering) duration from the start until the play-time of chunk $i$ is $\sum_{k=1}^i\alpha(k)$, where, as mentioned earlier, $\alpha(i)$ is the stall duration between the play of chunks $i-1$ and $i$. 
%, and $\gamma(i)$ is the the total stall duration before the $i$-th chunk including the startup delay ($\gamma(i)=S+\sum_{i^{\prime\prime}=1}^{i-1}\alpha(i^{\prime\prime})$). Remember, $S$ is the predefined startup delay. 
%Chunk $i$  starts playing at time given by  %$(i-1)L+\gamma(i)+\alpha(i)$. i.e, 
%$deadline(i)=S+(i-1)L+\sum_{k=1}^i\alpha(k)$.
%$i \in [i^\prime,C]$

Now, we describe our problem formulation given available bandwidth prediction for $W$ chunks ahead.
Let's assume that the current $W$ chunks are the chunks from $i^\prime$ to $C$ where $C=i^\prime+W-1$, and the current time is $j^\prime$ (time from the start of the download). We refer to the total stall duration before this segment plus the start up delay by $s$, where $s=S+\sum_{k=1}^{i^\prime-1}\alpha(k)$. Let $d(i)$ be the total stall duration encountered before the playback of the $i$-th chunk in the playback time of chunks in the segment $\{i^\prime, C\}$. i.e, $d(i_1) > d(i_0),\text{ if }i_1 > i_0$. With these settings, we formulate an optimization problem that \emph{(i)} minimizes the stall duration, \emph{(ii)} maximizes the  playback rate of the video accounting for diminishing returns with increase in chunk quality levels, and \emph{(iii)} minimizes the quality changes between the neighboring chunks to ensure the perceived quality is smooth. We give a higher priority to \emph{(i)} as compared to \emph{(ii)} and \emph{(iii)}, since stalls cause more quality-of-experience (QoE) degradation compared to playing back at a lower quality~\cite{MPC}. In particular, we consider an objective function that prefers pushing all the chunks to the $n$-th quality level over any other choice that might push the quality of some chunks to levels that are $> n$ with the cost of dropping the quality of some other chunks to levels $< n$.

  To account for the above objectives, we choose the objective function to be: $\big(\sum_{n=1}^{N}\beta^n\sum_{i=i^\prime}^{C}  I_{n,i}\big)-\lambda \dot d(C)$, where, $0<\beta<1$, and  $\lambda>>1$. The first part of the objective is a weighted sum of the level decision variables. More precisely it is a sum of the chunks obtained at least at the lowest quality plus $\beta$ times the number of chunks obtained at least at the second quality level, and so on.  %Therefore, pushing a chunk to the $n$-th quality level will achieve a utility that is $\beta$ times the utility achieved for the $(n-1)$-th level. %The proposed formulation maximizes a weighted sum of the quality level decision variables in which the contribution of increasing a chunk's quality from $n$ to $(n+1)$-th quality level achieve an objective of $0<\beta<1$ times the objective of increasing the smallest size chunk from $(n-1)$-th to $n$-th quality level. 
More precisely, $\beta$ should satisfy the following condistion:
\begin{equation}
%C\sum_{k=1}^{N-a} \gamma^{(k)} r_{a+k} < r_a, \  \text{  for } a= 0, \cdots, N. \label{basic_gamma_1}
\sum_{i=i'}^C\sum_{k=n+1}^{N} \beta^{k}  <  \beta^{n} , \  \text{  for } n= 0, \cdots, N. \label{basic_gamma_1}
\end{equation}
 This choice of $\beta$ implies diminishing returns with increasing quality levels. Fetching a chunk at quality $n$ has more utility as compared to improving quality of the rest of the chunks beyond $n$. Thus, the use of $\beta$ helps maximizing the minimum quality level among all chunks. This choice of $\beta$ will not increase the quality of some chunks beyond the $n$-th quality level at the cost of dropping the quality of one or more chunks bellow the $n$-th quality level. The second term in the objective function is the the total stall duration. We assume $\lambda>>1$. Therefore, avoidance of the stalls is given the highest priority. Due to these weights, the proposed algorithm will avoid stalls as the first priority and will not use the available bandwidth to increase the quality of some chunks at the expense of minimum chunk quality since lower levels are more preferable mirroring concave QoE function of the chunk rate.  The optimization problem for the window $[i^\prime, C]$, where $C=i^\prime+W-1$ can be formulated as follows, where $I_{n,i}$ is the decision variable for the $n$-th quality level for chunk $i$.

\begin{eqnarray}
&&\textbf{Maximize: } \Bigg(\sum_{n=0}^{N}\big(\beta^n\sum_{i=i^\prime}^{C}  I_{n,i}\big)-\lambda\cdot d(C)\Bigg)
\label{equ:eq1}
\end{eqnarray}
subject to
%\begin{eqnarray}
%\sum_{j=j^\prime}^{(i-1)L+s} z_0(i,j) = Y_{0,i},\forall i%  \quad  i = i^\prime, \cdots, C
%\label{equ:c1eq1}
%\end{eqnarray}
\begin{eqnarray}
I_{0,i} = 1,\forall i%  \quad  i = i^\prime, \cdots, C
\label{equ:c1eq1}
\end{eqnarray}

\begin{eqnarray}
\sum_{j=j^\prime}^{deadline(i)} z_n(i,j) = Z_{n,i},\forall i,n%\quad  i = i^\prime, \cdots, C, n = 1, \cdots, N
\label{equ:c2eq1}
\end{eqnarray}
\begin{eqnarray}
 I_{n,i}\le I_{n-1,i},\quad  \forall i,  n>1
\label{equ:c2_2eq11}
\end{eqnarray}
\begin{eqnarray}
 Z_{n,i}= I_{n,i}\cdot Y_{n,i},\quad  \forall i,  n
\label{equ:c2_2eq111}
\end{eqnarray}
%\vspace{-.1in}
\begin{eqnarray}
\sum_{n=0}^N\sum_{i=i^\prime}^{C} z_n(i,j)  \leq B(j), \forall j% \  \    j=j^\prime, \cdots, deadline(C)
\label{equ:c3eq1}
\end{eqnarray}
\begin{eqnarray}
\sum_{i, deadline(i) > t} {\bf I}\Bigg(\sum_{j=1}^t\bigg(\sum_{n=0}^Nz_n(i,j)\bigg)> 0\Bigg) L \leq B_m \   \forall t
\label{equ:c4eq1}
\end{eqnarray}
\begin{equation}
z_n(i,j) \geq 0\ ,  \forall i,j% j=1, \cdots, deadline(C)
\label{equ:c5eq1}
\end{equation}
\begin{equation}
z_n(i,j) = 0\   \forall i, j > deadline(i)
\label{equ:c6eq1}
\end{equation}
\begin{equation}
I_{n,i} \in \{0, 1\}, \forall i,n% \quad  \forall i = 1, \cdots, C, \text{ and } \forall n = 1, \cdots, N
\label{equ:c7eq1}
\end{equation}
\begin{equation}
d(i)\ge 0, deadline(i) = S+(i-1)L+\sum_{k=1}^{i^\prime-1}\alpha(k)+d(i)
\label{equ:c8eq1}
\end{equation}

\begin{eqnarray}
\text{Variables:}&& z_n(i,j), I_{n,i}, Z_{n,i}, d(i) \ \ \  \forall   i = i^\prime, \cdots, C,  \nonumber \\
&&j = j^\prime, \cdots, deadline(C), \  n = 0, \cdots, N \nonumber
\end{eqnarray}

%\begin{equation}
%X(i) \in {\mathcal X}_i \quad  \forall i = i^\prime, \cdots, C
%\label{equ:c7eq1}
%\end{equation}

%\begin{eqnarray}
%\text{Variables:}&& x(i,j), X(i) \ \ \  \forall   i = 1, \cdots, C,  \nonumber \\
%&&j = 1, \cdots, deadline(C), \nonumber
%\end{eqnarray}
Constraint \eqref{equ:c1eq1} ensures that every chunk is fetched at least at lowest quality, \ie there are no skips. Constraint \eqref{equ:c2eq1} defines the total amount fetched for a chunk $i$. Constraint \eqref{equ:c2_2eq11} ensures that if a chunk is not a candidate to $(n-1)$-th level, it won't be a candidate to the $n$-th level, so it should not be considered for the $n$-th quality level. Constraint \eqref{equ:c2_2eq111} enforces $Z_{n,i}$ to be in $\{0, Y_{n,i}\}$.
\eqref{equ:c3eq1} imposes the available bandwidth constraint at each time slot $j$, and \eqref{equ:c4eq1} imposes the playback buffer constraint so that the total playback in the buffer at any time does not exceed the buffer capacity $B_m$, where ${\bf I}({\bf \cdot})$ is an indicator function. Therefore, as far as a chunk is partially downloaded, the buffer duration is increased by a duration of a chunk. 
Constraint \eqref{equ:c5eq1} imposes the non-negativity of chunk download, and \eqref{equ:c6eq1} imposes the deadline constraint, so no chunk is fetched after its deadline.  \eqref{equ:c7eq1} enforces the decision variables $Z_{n,i}$ to belong to the discrete set $\{0,Y_{n,i}\}$. Finally, Constraint \eqref{equ:c8eq1} states that stall duration before any chunk is non-negative and defines the deadline of any chunk $i$.

The problem \eqref{equ:eq1}-\eqref{equ:c8eq1} is a non-convex optimization problem since the feasible set (the set of the decision variables) is discrete, and the problem contains non-convex constraint (\ref{equ:c4eq1}). However, we will show later that when the video is CBR encoded per quality level ($X_{n,i}=X_n, \forall i$, where $X_n$ is a constant for the $n$th quality level), the proposed algorithm achieves the optimal solution to the problem \eqref{equ:eq1}-\eqref{equ:c8eq1}.

%% file: bwPrediction.tex
%\section{Bandwidth Prediction}
%\label{bwPrediction}

{\bf Bandwidth Prediction:} We note that the proposed formulation depends strongly on the available bandwidth prediction. In practice, perfect prediction cannot be non-causally available. There are multiple ways to obtain the prediction, including a crowd-sourced method to obtain historical data \cite{riiser2013commute,GTube}. Another approach may be to use a function of the past data rates obtained as a predictor of the future, as an example is to compute the exponential weighted moving average or the harmonic mean of the download time of the past $\eta$ chunks to predict the future available bandwidth ~\cite{chen2016msplayer}. 
%
%The weighted moving average (WMA) of the $(t+1)$th segment is defined as:
%\begin{equation}
%\hat{B}(t + 1) = \alpha \cdot \hat{B}(t) + (1 - \alpha) \cdot B(t)
%\end{equation}
The weighted moving average smooths out the fluctuation of the available bandwidth measurements
and the weight is given to the latest data.
The harmonic mean method has been shown to mitigate the impact of large outliers due to network variations \cite{jiang2012improving}.
Therefore, in this paper, we consider using the harmonic mean of the download time of the last $\eta$ chunks to predict the future available bandwidth. Given available bandwidth measurements of the last past $\eta$ chunks, the predicted available  bandwidth of the next segment is calculated as 
\begin{equation}
\hat{B}(n+1)=\frac{\eta}{\frac{1}{\sum_{n-\eta}^{n}B(t)}}.
\end{equation}

%% file: algorithm.tex
\section{FastScan Algorithm}
\label{sec:alg}

We describe the algorithm for adaptively streaming videos with qualities being decided based on the predicted available bandwidth, chunk deadlines, and buffer occupancy. 
The objective is to determine up to which quality level we need to fetch each chunk for the next window of chunks, such that the  stall duration is minimized, and the proposed concave QoE metric in playback rate is maximized. 

%average playback bitrate is maximized and the quality switching rate is minimized as the second priority.
%\input{avcAlgo}

We will mainly describe the algorithm for a prediction window of $W$ chunks. However, the available bandwidth prediction is updated after the download of each chunk using the harmonic mean based prediction technique. Thus, the quality decisions will be updated after each chunk's download based on the updated bandwidth prediction. 

\begin{figure}[h]
\label{algorithm 1}
		%\vspace{-.2in}
		\begin{minipage}{\linewidth}
			\begin{algorithm}[H]
				\small
				\begin{algorithmic}[1]
			\STATE {\bf Input:} ; $X_n(i)\forall i,n$, $N$, $s$, $B_m$, $i^\prime$, $j^\prime$, $W$, $B(j)\forall j$, 
			\STATE {\bf Output:} $X(i) \forall i$: The maximum size in which chunk $i$ can be fetched, $I_n$: set contains the indices of the chunks that can be fetched up to $n$th quality level.
					
    \STATE {\bf Initialization:}
    \STATE $C=i^\prime+W-1$
    \STATE $X(i)=0$, $d(i)=0$, $deadline(i) = s+(i-1)L$, $\forall i$
   \STATE $c(j)=\sum_{j^\prime=1}^{j}B(j^\prime)$ available bandwidth up to time $j$, $\forall j$
   \STATE $bf(j)=0$: buffer length at time $j$
   \STATE $t(i)=0, \forall i$, first time slot in which chunk $i$ can be fetched
   %\STATE $a(i)=0, \forall i$, lower layer decision of fetched amount of chunk $i$ at its lower deadline time $t(i)$
   \STATE $e(j)=B(j), \forall j$,  remaining  available bandwidth at time j after all non skipped chunk are fetched %according to lower layer size decisions 
   \STATE {\bf Lowest Quality level decision, $n=0$:}
    \STATE $d=$ {\bf level-0-Forward}$(B, X, i^\prime, C,L,d, deadline,$ $B_m, bf, j^\prime)$
    \STATE $deadline(i)=(i-1)L+s+d(C)$
     \STATE $d=$ {\bf level-0-Backward}($B$,$X_0$,$x$,$i^\prime$,$C$,$L$,$d$,$deadline$,$B_m$,$bf$,$j^\prime$)
      \STATE $deadline(i)=(i-1)L+s+d(i)$
  \STATE {\bf Higher Quality Levels' decisions:}
   \STATE {\bf For every quality level $n > 0$}
   \STATE $[t,a,e] =$ {\bf level-n-forward}($B$,$X$,$i^\prime$,$C$,$L$,$deadline$,$B_m$,$bf$,$j^\prime$,$I_0$)
  \STATE [$X$,$I_n$]$=${\bf level-n-backward}($B$,$X$,$X_{n}$,$i^\prime$,$C$,$L$,$deadline$, $B_m$,$bf$,$j^\prime$,$t$,$c$,$a$,$e$)
   				\end{algorithmic}
				\caption{FastScan Algorithm}\label{algo:DABP}
			\end{algorithm}
		\end{minipage}
		\vspace{-.1in}
	\end{figure}

%The key idea of the FastScan algorithm is to start from seeing if the chunks can be obtained at the lowest quality, and determine stalls to obtain the chunks at the lowest quality. This will provide the minimum stall duration. Having calculated that, we then go over the next level to see with the current stall duration, which chunks can now be improved to be obtained at the next quality level. Thus, we will try to bin-pack higher quality possibilities one by one. Thus, there is a strict preference to improve the lowest quality level of the chunks. Since the bin packing will continue through the levels, the wastage of bandwidth could happen if the quality of no chunk could be improved due to  deadline and buffer constraints. 

\begin{figure}
\centering
%\vspace{-0.25in}
\includegraphics[trim=0.1in 0.8in 0.1in 1.2in, clip,width=0.48\textwidth]{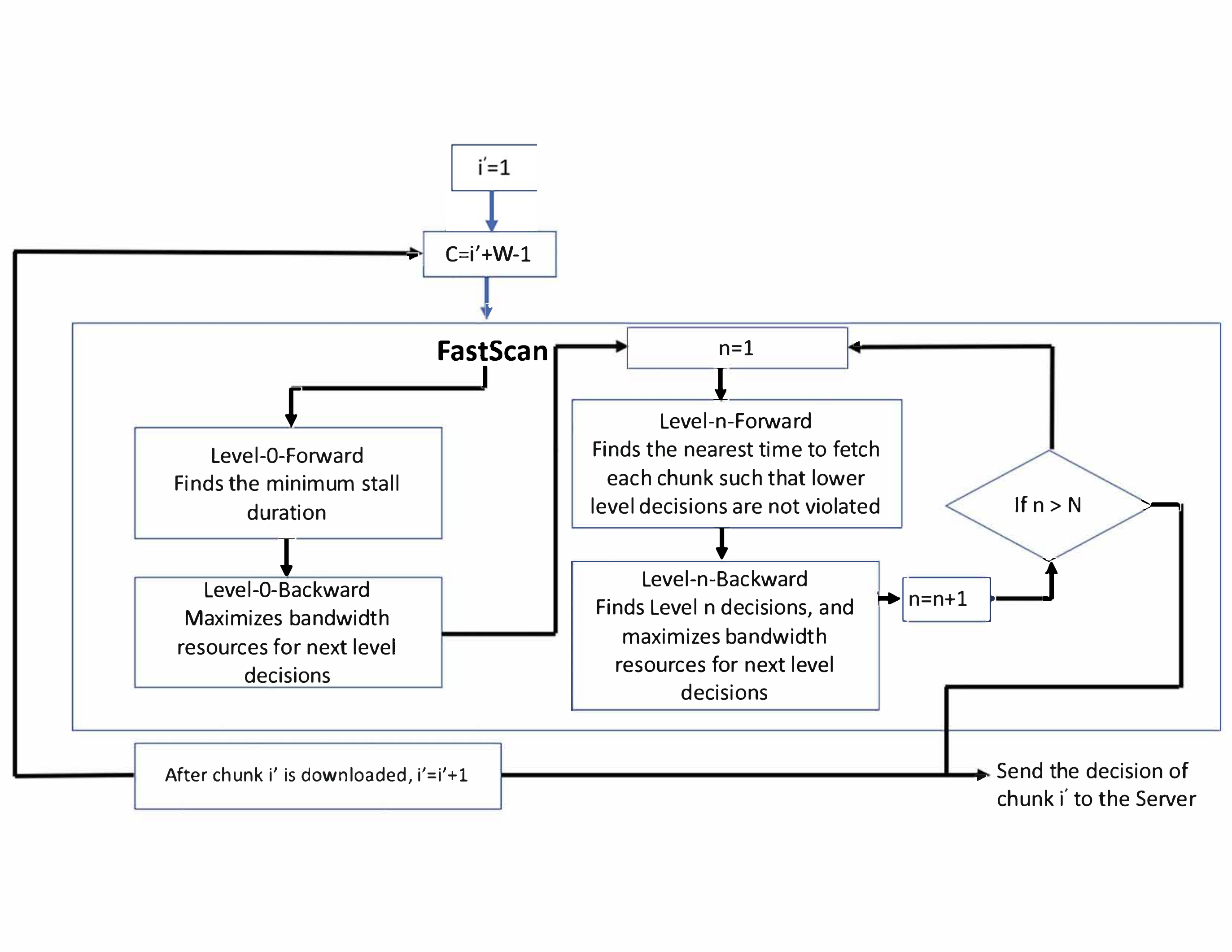}	
\vspace{-0.2in}
 \caption{Flowchart Diagram of FastScan Algorithm}
 \label{fig : fastScan_Algo}
 \vspace{-.2in}
 \end{figure}

As shown in Fig.~\ref{fig : fastScan_Algo}, FastScan algorithm is called after the download of every chunk to make the quality decisions of the next $W$ chunks starting from chunk $i^\prime$. Therefore, the algorithm is called in sliding window manner in which previous decisions of some chunks may change on fly plus considering one new chunk in every call. Before, we  describe the algorithm in detail, we introduce the high level idea of the algorithm as explained in Fig.~\ref{fig : fastScan_Algo}. At a high level, FastScan performs forward followed by backward scans at each quality level. The scans simulate fetching chunks in forward and reverse order, respectively. The Level-0 forward and backward scan's objective is to find the minimum stall duration and maximize the resources of the next level decision as it will be explained in more detail later. However, starting from level 1, the job of every Level-n forward and backward scans is to maximize the number of candidate chunks to the $n$-th quality level without violating the lower level decisions.

The algorithm is summarized in Algorithm~\ref{algo:DABP}, and it works as follows. Given the size of the lowest quality level for every chunk, chunks' deadlines, predicted available bandwidth, and buffer size, FastScan performs a forward scan followed by a backward scan by calling Level-0 Forward and Level-0 Backward algorithms (Algorithms~\ref{algo:forward0} and \ref{algo:backward0}) assuming that all chunks need to be fetched at the lowest quality level ($0$th quality level). Based on the first pass of the forward and backward algorithms, decisions of the stall duration, stall pattern and the final deadline of every chunk such that the stall duration is minimized and the bandwidth resources to every chunk are maximized are found. Consequently, the Level-$n$ Forward and Backward ($n=1,\cdots,N$) algorithms are performed in order to find the candidate chunks to the $n$th quality level. The final quality decision for every chunk in the current window of chunks will be the return of the backward algorithm's call of the highest quality level. The algorithm is re-run after the download of every chunk, so it keeps adjusting for the prediction error and changing decisions if necessary on the fly. In order to reduce the complexity, the algorithm uses the cumulative predicted available bandwidth of the $j$th second ($c(j)$, line 5), so the algorithm avoids having double loops summing the available bandwidth over multiple time slots when making the decision per chunk. 

Level-0-forward algorithm finds the minimum stall time such that all chunks are fetched at the lowest quality level, then the algorithm assumes that all stalls can be brought to the beginning of the current segment by setting the deadline of every chunk in this segment to the value defined by line 11. Level-0-Backward algorithm (line 12) runs after that, and its job is to check if all stalls can indeed be brought to the beginning of this window of chunks, if not, the algorithm will find the minimum number of later stalls. The goal of the backward algorithm is to maximize the number of candidate chunks to the next level since if all stalls can be shifted to their earliest, since in this case all chunks have higher chance to increase their qualities. Therefore, the deadline that is found after running Level-0-backward algorithm (line 13) is final. Finally, the Level-n Forward and Backward algorithms are run per level to find the chunks that can still be fetched at higher quality levels.  We will now describe the  forward and backward algorithms.

 %checks the possibility of fetching a chunk at any of the quality levels sequentially. First step is checking if all chunks can be fetched at lowest quality  works from the lowest to the highest quality level, and processes each quality level for the chunks that are candidates for the previous one. The algorithm performs forward and backward scans (explained bellow) in sequence starting from the lowest quality level ({\bf level-0})(lines 7-9, and 13-14). %we described is ``{\em No-Skip Deadline and Buffer Aware Bin Packing Algorithm}" (Algorithm 5). An example run of the algorithm is described in Appendix \ref{noskipex}.
 
\begin{figure}[hbtp]
		%\vspace{-.6in}
		\begin{minipage}{\linewidth}
			\begin{algorithm}[H]
				\small
				\begin{algorithmic}[1]
				\STATE {\bf Output:} $d(i)$: stall time of chunk $i$.

 \STATE $ j=j^\prime$, $i=i^\prime$
    \WHILE {chunk $C$ is not fetched yet}
      		\IF{$(i > 1$ and $d(i) < d(i-1))$}
			\STATE $d(i)=d(i-1)$, $deadline(i)=(i-1)L+s+d(i)$
		\ENDIF
		\STATE{\bf if} $(bf(j) = Bm)$, {\bf then} $j=j+1$
                 \STATE $fetched=min(B(j), X(i))$
                 \STATE $B(j)=B(j)-fetched$, $X(i)=X(i)-fetched$
                 \STATE {\bf if} $X(i) > 0$, {\bf then} $bf(j)=bf(j)+L$
                 \IF{$X(i)=0$ and $j <= deadline(i)$}
                      \STATE i=i+1
                \ELSIF{$X(i)=0$ and $j >deadline(i) $}
                \STATE $d(i)=d(i)+j-deadline(i)$
                     % \STATE $bf(t(i):deadline(i))=bf(t(i):deadline(i))+L$
                      \STATE $i=i+1$
                \ENDIF

                 \STATE {\bf if} $B(j)=0$, {\bf then} $j=j+1$
                   
                      \ENDWHILE

				\end{algorithmic}
				\caption{{\bf Level-0} Forward Algorithm}\label{algo:forward0}
			\end{algorithm}
		\end{minipage}
%		\vspace{-.2in}
	\end{figure}

		\begin{figure}[hbtp]
		\vspace{-.1in}
		\begin{minipage}{\linewidth}
			\begin{algorithm}[H]
				\small
				\begin{algorithmic}[1]
					\STATE {\bf Output:} $d_f(i)$: final stall duration of chunk $i$.
					
    \STATE {\bf Initilization:}
    \STATE $i=C$, $ j=deadline(C)$
\STATE $d_f(i)=d(i)$, and $deadline(i)=(i-1)L+s+d_f(i)$ $\forall i$
     \WHILE {($j >( j^\prime-1)$ and $i > (i^\prime-1)$)}
     \IF{$(i < C)$}
     	\STATE $d_f(i)=d(i)-(d(i+1)-d_f(i+1))$ 
	\STATE $deadline(i)=(i-1)L+s+d_f(i)$
     \ENDIF
        \IF{$j <= deadline(i)$}
        			\IF{($bf(deadline(i)) = B_m)$}
				\STATE $d_f(i)=d(i)-1$
				\STATE $j=j-1$
               		\ELSE
               		\STATE $fetched=min(B(j),X_0(i))$
               		\STATE $X_0(i)=X_0(i)+fetched$, $B(j)=B(j)-fetched$
                 	\STATE {\bf if}$(X(i) > 0)$, {\bf then} $bf(j)=bf(j)+L$
                		\STATE {\bf if}$(X_0(i) = 0)$, {\bf then} $i=i-1$
                	 	\STATE {\bf if}$(B(j) = 0)$, {\bf then} $j=j-1$
		\ENDIF
           \ELSE
           	\STATE $j=j-1$
        \ENDIF
\ENDWHILE

				\end{algorithmic}
				\caption{{\bf Level-0} Backward Algorithm}\label{algo:backward0}
			\end{algorithm}
		\end{minipage}
		\vspace{-.2in}
	\end{figure}

	\if 0
	\begin{figure}[hbtp]
		%\vspace{-.6in}
		\begin{minipage}{\linewidth}
			\begin{algorithm}[H]
				\small
				\begin{algorithmic}[1]
					\STATE {\bf Output:} $d_f(i)$: final stall duration of chunk $i$.
					
    \STATE {\bf Initilization:}
    \STATE $i=C$, $ j=deadline(C)$

     \WHILE {($j >( j^\prime-1)$ and $i > (i^\prime-1)$)}
     \IF{$(i < C)$}
     	\STATE $d_f(i)=d(i)-(d(i-1)-d_f(i-1))$ 
	\STATE $deadline(i)=(i-1)L+s+d_f(i)$
     \ENDIF
        \IF{$j <= deadline(i)$}
        			\IF{($bf(deadline(i)) = B_m)$}
				{\bf then} $d_f(i)=d(i)-1$
				\STATE $j=j-1$
               		\ELSE
               		\STATE $fetched=min(B(j),X_0(i))$
               		\STATE $X_0(i)=X_0(i)+fetched$, $B(j)=B(j)-fetched$
                 	\STATE {\bf if}$(X(i) > 0)$, {\bf then} $bf(j)=bf(j)+L$
                		\STATE {\bf if}$(X_0(i) = 0)$, {\bf then} $i=i-1$
                	 	\STATE {\bf if}$(B(j) = 0)$, {\bf then} $j=j-1$
		\ENDIF
           \ELSE
           	\STATE $j=j-1$
        \ENDIF
\ENDWHILE

				\end{algorithmic}
				\caption{{\bf Level-0} Backward Algorithm }
			\end{algorithm}
		\end{minipage}
%		\vspace{-.2in}
	\end{figure}
\fi
\vspace{0.1in}
{\bf Level-0 Forward algorithm}: %Given the predicted bandwidth, the buffer size, and the prediction window size $W$, the algorithm simulates fetching the next $W$ chunks at the lowest quality level and determines the minimum stall duration of this segment since that is the first priority.  
To explain Level-0 Forward algorithm, let's assume that the current $W$ chunks that need to be fetched are the chunks from $i^\prime$ to $C$ where $C=i^\prime+W-1$.  The algorithm simulates fetching these chunks in order at the lowest quality level with an assumption that the stall duration of each chunk in this segment is zero, \ie $d(i^\prime)=\cdots = d(C)=0$, $\forall i \in \{i^\prime,C\}$. Therefore, the initial deadline of every chunk $i$ such that $i^\prime \leq i \leq C$ is $(i-1)L+s$, where $s=S+\sum_{k=1}^{i^\prime-1}\alpha(k)$ is the total stall duration before the current segment. Chunks are fetched in order and if a chunk $i$ can't be completely downloaded before  its initial deadline, the additional time spent in fetching this chunk is added to $d(i^{\prime\prime})$ for every $i \leq i^{\prime\prime} \leq C$ (lines 13-16) since there has to be an additional stall in order to fetch these chunks. Therefore,  level-0 Forward algorithm finds the total stall duration of this segment of chunks and the deadline of the last chunk ($d(C)$, and $deadline(C)$) such that all chunks are guaranteed to be fetched at least at the lowest quality level. %Finally, Level-0 Forward algorithm is repeated after the download of every chunk.
%$d(i)$ is defined as $\sum_{k=i^\prime}^i\alpha(k)$
Two things to note, first, Level-0-forward algorithm gives a guarantee that all chunks from $i^\prime$ to $C$ can be fetched before the time slot $deadline(C)$ at least at the lowest quality level with the minimum stall duration according to the predicted available bandwidth. Second, forward algorithm may lead to stalls in the middle of the segment which can hurt the performance by having less opportunity to fetch later chunks at higher quality levels. To avoid this situation, we run the Level-0 Backward algorithm right after the Level-0 forward algorithm to move stalls earlier helping more chunks to possibly be fetched at higher quality. 

{\bf Level-0 Backward algorithm}: %This algorithm starts from the last time slot ($deadline(C)$), and simulate fetching chunks closer to their deadlines in order to move stalls to the start of the segment (before chunk $i^\prime$). 
Given the predicted available bandwidth, the forward algorithm decisions, and the buffer size, Level-0 Backward algorithm (Algorithm 3) simulates fetching chunks at {\bf level-0} quality  in reverse order starting at the deadline of the last chunk ($j=deadline(C)= (C-1)L+s+d(C)$). Therefore, the Level-0 Backward algorithm simulates fetching chunks close to their deadlines. The backward algorithm's objective is to move stalls to their earliest time, \ie as early as possible. In fact, if the buffer is infinite, all stalls can be brought to the beginning of the current segment, but if the buffer is finite, that may not always be possible. There will be one scenario that leads to changing the initial deadline and having a stall that is not brought to the very beginning of the current segment which is the buffer constraint violation (lines 11-13), \ie If fetching a chunk $i$ results in a buffer constraint violation, the algorithm decrements the deadline of chunk $i$ by 1 and checks if the violation can be removed. This decrement will continue until the buffer constraint is no longer violated. The aim of the Level-0 Backward algorithm is to provide the deadlines of the different video chunks such that the fetching of the video minimizes the stall duration. Further, Level-0 Backward algorithm aims to bring all the stalls to their earliest possible time, since this provides all chunks  more time to increase their qualities without violating $deadline(C)$. Therefore, later chunks have a higher chance of increasing their quality.

Once {\bf Level-0} decisions are found, the {\bf Level-n} forward, and the {\bf Level-n} backward algorithms (Algorithm 4, and 5) are run per level. The key difference of {\bf Level-n} forward and {\bf Level-n} backward algorithms as compared to {\bf Level-0} forward and backward algorithms is that the deadlines are fixed and can't be changed. Therefore, if a chunk can't be fetched at $n$th quality level with its current deadline, it is not considered as a candidate to that quality level.

%$n$th layer forward scan objective is to find lower deadline per chunk in which the chunk can't be fetched before; otherwise previous level decisions may become violated.

{\bf Level-n} forward algorithm simulates fetching chunks in order, where chunks are fetched according to the $(n-1)$th level decision. The main objective of {\bf Level-n} forward algorithm is to determine the lower deadline of every chunk $i$ in which chunk $i$ can't be fetched earlier ($t(i)$, lines (10-11)); otherwise, the lower level decisions are violated. As we will see next, the lower and upper deadlines of every chunk are the key components that {\bf Level-n} backward algorithm use to make a decision of which chunk can be fetched at the $n$th quality level.

% such that the number of chunks fetched at $n$th quality level is maximized without violating the lower quality level decisions. i.e. {\bf level-n} forward algorithm finds the chunks that can be posted to next quality level without creating un-necessary and more than one quality level switching between neighboring chunks preserves the smoothness of the quality level 

%with one change which is assuming that chunks that were candidates to $(n-1)$th quality level are fetched at $n$th quality level. i.e. $n$th level is considered only for chunks that were candidates for $(n-1)$th quality level

{\bf Level-n} backward algorithm takes as input the lower and the upper deadlines of every chunk and finds the set of chunks that can be pushed to the $n$th quality level. If pushing a chunk to $n$th quality level is not possible without violating its lower and upper deadlines, the chunk will not be a candidate to the $n$th quality level. \textit{Therefore, we clearly see that the algorithm prioritizes pushing the maximum number of chunks to the $(n-1)$-th quality level over fetching some at the $n$-th quality level with the cost of reducing the quality of some others below the $(n-1)$th quality level. This will prioritize horizontal over vertical scanning which reduces the quality switching that is spanning more than one quality-level.}
%{\bf level-n Forward algorithm}: (Algorithm 3) is run to simulate fetching chunks in order with quality decisions up to $(n-1)$th level and provides lower deadlines of chunks for the level-n backward run. Then, forward  scan finds the lower deadline of every chunk, so the lower level decisions are not violated.

 %Running backward algorithm at $n$th level is the most tricky part since. 
 %Given the predicted available bandwidth, the lower and the upper deadlines of every chunk, and the buffer size, the algorithm simulates fetching the chunks in reverse order. The algorithm initially increases the size of every chunk that was a candidate to the $(n-1)$-th quality level to the $n$-th quality level size. For every chunk $i$, the backward algorithm checks the remaining available bandwidth between its lower and the upper deadlines ($t(i)$, and $deadline(i)$) after all chunks $> i$ are fetched and the buffer. 

 The backward algorithm considers if the chunk can be a candidate for the $n$th quality level. In order to decide this, the algorithm determines if there is enough available bandwidth and the buffer is not full (line 17-21). If chunk $i$ was a candidate to the $(n-1)$-th quality level and is not selected to be fetched at the $n$-th quality level. If the chunk was a candidate to the $(n-1)$-th quality level and is not selected to be fetched at the $n$th quality level, this could be because of one of the two scenarios mentioned next. The first is the violation of the {\bf buffer} capacity, where the chunk could not be placed in the buffer before its deadline (line 7). The second is the {\bf available bandwidth} constraint violation where the remaining available bandwidth is not enough for downloading the chunk at the $n$-th quality level (lines 14-16).

 %This scenario also means that the chunk could not be fetched by its deadline, so it can also be called deadline violation.

For buffer capacity violation, we first note that there could be a chunk  $i^{\prime\prime} > i$, which if it is not fetched at the $n$-th quality level, chunk $i$ could have been an $n$-th level candidate. However, {\bf Level-n} backward algorithm decides to not consider $i$ (line 7). We note that since there is a buffer capacity violation, one of the chunks $\geq i$ must be skipped (not considered for the $n$-th level). The reason for choosing to not consider (skip) chunk $i$ rather than a later one is that $i$ is the closest one to its deadline. Therefore, $i$ is not a better candidate to the next quality level than any of the later chunks. 

In the second case of deadline/available bandwidth violation, {\bf Level-n}  backward algorithm decides to skip chunks up to $i$ since there is not enough available bandwidth. As before, an equal number of chunks need to be skipped (for CBR encoding), skipping earlier ones is better because it helps increasing the potential of getting later chunks at higher quality levels.

Finally, giving higher priority to later chunks reduces the effect of the inaccurate available bandwidth estimation. In other words, by giving priority to later chunks, if the real available bandwidth turns to be lower than the predicted one, the prediction error effect is minimized. Moreover, to handle prediction error, a lower buffer threshold can be set, so if the buffer is running lower than this threshold, the level decision that was made by our streaming algorithm for the next chunk is reduced by 1 quality level (except if the chunk is already at lowest quality).

\begin{figure}[hbtp]
		\vspace{-.2in}
		\begin{minipage}{\linewidth}
			\begin{algorithm}[H]
				\small
				\begin{algorithmic}[1]
				%\STATE {\bf Input:} $B, X, C, deadline, Bm, bf,I_0$
				\STATE {\bf Output:} $t(i)$, $a(i)$, $e(j)$

   \STATE $ j=1$, $k=1$
    \WHILE {$j \leq deadline(C)$ and $k \leq max(I_0)$ (last chunk to fetch)}
	\STATE $i=I(k)$
          \STATE {\bf if }{$i=0$} {\bf then } $k=k+1$
            \IF{$j \leq deadline(i)$}
             \STATE {\bf if } {$(bf(j) = B_m)$} {\bf then }  $j=j+1$
                 \STATE $fetched=min(B(j), X(i))$
                \IF{$j$ is the first time chunk $i$ is fetched}
                 \STATE $t(i)=j$, $a(i)=fetched$ 
                    \ENDIF

                 \STATE $B(j)=B(j)-fetched$, $e(j)=B(j)$, $X(i)=X(i)-fetched$

                   \STATE {\bf if } {$X(i) > 0$} {\bf then }  $bf(j)=bf(j)+L$% and $j \geq t(i)$}
                  \STATE {\bf if } {$X(i)=0$} {\bf then } $k=k+1$

                   \STATE {\bf if } {$B(j)=0$} {\bf then } $j=j+1$
               \ELSE
                \STATE $k=k+1$
            \ENDIF

   \ENDWHILE

				\end{algorithmic}
				\caption{{\bf Level-n} Forward Algorithm}\label{algo:forwardn}
			\end{algorithm}
		\end{minipage}
\vspace{-.1in}
	\end{figure}

\begin{figure}
		%\vspace{-.6in}
		\begin{minipage}{\linewidth}
			\begin{algorithm}[H]
				\small
				\begin{algorithmic}[1]
				%\STATE {\bf Input:} $B, X,X_0, C, L, deadline, B_m,bf,t,r,a,e$					\STATE {\bf Output:} $X(i)$, $I_n$
    \STATE {\bf Initialization:}
    \STATE $i=C$, $ j=deadline(C)$
     \WHILE {($j > 0$ and $i > 0$)}
        \IF{$j <= deadline(i)$}
        \STATE {\bf if} {$(bf(deadline(i)) = B_m)$} {\bf then } $i=i-1$
              % \STATE continue
%         \ENDIF

                \IF{$j$ is the first time to fetch chunk $i$ from back}
                		\IF{$(t(i)=0)$}
				\STATE $rem1=c(j)-c(1)+e(1)$, $rem2=rem1$
				%\STATE $rem2=rem1$
			\ELSE
			\STATE $rem2=c(j)-c(t(i))$, $rem1=rem2+e(t(i))+a(i)$
				
			\ENDIF
                		\IF{$(rem1 < X_n(i))$}
				\STATE {\bf if} {$(X(i) > 0)$} {\bf then} $X_n(i)=X(i)$ {\bf else } $i=i-1$
			 \ELSE
		        		
                			\IF{$(rem2 < X_n(i))$ and $rem1 \geq X_n(i))$}
                				\STATE $e(t(i))=e(t(i))+rem1-X_n$
				\ENDIF
			        \STATE $X(i)=X_n(i)$, $I_n \leftarrow I_n\cup i$
            		\ENDIF
          	\ENDIF
	
            \STATE $fetched=min(B(j), X_n(i))$ 
            \STATE $B(j)=B(j)-fetched$, $X_n(i)=X_n(i)-fetched$
            \STATE  {\bf if} {$(X_n(i) > 0)$} {\bf then } $bf(j)=bf(j)+L$
            	
            \STATE {\bf if } {$(X_n(i)=0)$} {\bf then } $i=i-1$
	
	   \STATE {\bf if } {$(B(j)=0)$} {\bf then }  $j=j-1$            	
            \ELSE
           \STATE $j=j-1$
        \ENDIF
\ENDWHILE
				\end{algorithmic}
				\caption{{\bf Level-n} Backward Algorithm }\label{algo:backwardn}
			\end{algorithm}
		\end{minipage}
		\vspace{-.2in}
	\end{figure}
	
{\bf Complexity Analysis}: We note that the initialization step sums the variables over time, and is thus O($W$) complexity. The backward and forward algorithms are run once per quality level. For each run of backward/forward algorithm, there is a while loop, and within the loop, the complexity is O(1). Since the while loop runs at most $W+deadline(i^\prime+W-1)+1$ times, the overall complexity of the proposed algorithm is O($NW$). In order to decrease the complexity, cumulative available bandwidth for every time slot $j$, $c(j)$ is used to avoid summing over the available bandwidth in the backward and the forward loops.

{\bf Optimality of FastScan Algorithm for CBR encoded videos: } We note that the proposed algorithm is a variant of the algorithm proposed for SVC encoded videos in \cite{AnisTONSingle}. Adapting the results in \cite{AnisTONSingle}, the optimality of the proposed algorithm in solving \eqref{equ:eq1}-\eqref{equ:c8eq1} follows. Therefore, in the offline case in which the bandwidth is perfectly predicted for the whole period of the video, and $W=V$, the obtained quality is the offline optimal. The result is summarized in the following theorem, and the proof is omitted since it is an adaptation of the result in \cite{AnisTONSingle}.

\begin{theorem}%{Theorem}
Up to a given quality level $M, M \geq 0$, if $(I_{m,i}^*, d^*(i))$ are the  $m$-th quality level decision variable $m \leq M$ and stall duration of chunk $i$ that are found by using FastScan algorithm, and $I_{m,i}^\prime,d^\prime(i)$ are the decision variable and the stall duration that are found using any other feasible algorithm, then the following holds for any $0<\beta < 1$, satisfies (\ref{basic_gamma_1}).
\begin{equation}
 \sum_{m=0}^M \beta^m\sum_{i=i^\prime}^{C} I_{m,i}^\prime -\lambda d^\prime(C) \leq \sum_{m=0}^{M} \beta^m\sum_{i=i^\prime}^{C} I_{m,i}^*-\lambda  d^*(C).
\label{equ:thm}
\end{equation}
In other words, FastScan Algorithm finds the optimal solution to the optimization problem~(\ref{equ:eq1}-\ref{equ:c8eq1}) when $0<\beta < 1$,  (\ref{basic_gamma_1}) holds, and $\lambda \gg 1$.
\label{theorem: theorem1}
\end{theorem}

{\bf Discussions: } We note that the optimization problem in this paper is a combinatorial optimization problem, with discrete constraints. Many  known combinatorial optimization problems  are NP hard (\eg Knapsack problem). We also note that the combinatorial optimization problem is not necessarily NP hard. For example, matching problem is one of the combinatorial problem that is not NP-hard. The NP-hard Knapsack problem optimizes a linear function of integer variables with a single linear constraint. The proposed problem has multiple constraints, which intuitively makes it harder. However, the structure in the problem considered in this paper allows us to find optimal algorithm which has linear time complexity. We note that a straightforward greedy level-by-level optimization will not be optimal. This is because the decisions at the higher levels depend on the decisions at the lower levels. Thus, the decisions at the lower levels must leave large available bandwidth for fetching the incremental content of higher quality. The algorithm in this paper helps account for the connections for different layers and is optimal for the proposed problem. Thus, the diminishing returns in increasing quality levels make both a practical motivation and leads to a computationally simple optimal algorithm for CBR encoded videos.

%Some of the
%problems in this class are the Knapsack problem, Cutting
%stock problem, Bin packing problem, and Travelling salesman
%problem. These problems are all known to be NP hard. Very
%limited problems in this class of combinatorial optimization

%% file: implementation.tex
\section{Implementation}

In this section, we describe our implementation of the proposed
approach in the dash.js framework. Our implementation
is based on the dash.js master branch (v1.2.0 release) with the modifications proposed by \cite{MPC}.

\subsection{Setup}

Our choice of dash.js framework is motivated by the fact that dash.js is a reference open-source implementation for the MPEG-DASH standard and is actively supported by leading industry participants~\cite{MPC}.  The system architecture of dash.js is shown in Fig.~\ref{fig : dashjsSys}. Briefly, dash.js abstracts the high-level video streaming functionalities such as the rate adaptation logic on top of the low-level DASH standard related components. The main class that is responsible for the rate adaptation techniques is AbrController. This class contains the core bitrate adaptation logic. In the original dash.js implementation, a rule-based decision logic is implemented to find the bitrate. Specifically, DownloadRatioRule selects bitrate based on the download ratio (play time of last chunk divided by its download time). On the other hand, InsufficientBufferRule chooses bitrate depending on whether the buffer level has reached a lower limit recently to avoid re-buffers. Priorities are assigned to each rule to resolve conflicts and make final bitrate decisions \cite{MPC}. Further, we use the modifications to  dash.js that was proposed by \cite{MPC} which allow it to work with prediction based adaptation algorithms. For details on such modifications, the reader is referred to \cite{MPC}.

%Modifications and extensions added by \cite{MPC}:
%First modification is mandating the manifest to report chunk
%sizes, which can be used along the bandwidth prediction to make decisions on the quality of the next few chunks.
%
%Secondly, modify the original dash.js to make the chunk decisions at chunk boundary after downloading a chunk and before downloading the next one.
%
%Third, the original dash.js downloads multiple chunks in parallel
%even though chunks that are earlier in the video stream
%should ideally be prioritized. However, \cite{MPC} modified the bitrate decision
%and chunk download process in dash.js code by
%making two key changes to BufferController class:
%1) bitrate decisions are made at the start of each chunk, 2)
%chunk download is completely sequential, i.e., no concurrent
%downloads of multiple chunks are allowed. This allows
%a basic implementation framework which is consistent with
%our model and other proposed algorithms.
%
%
%Additional to these modifications , \cite{MPC} implemented different bitrate adaptation algorithms such as their proposed algorithm FastMPC, and some other buffer and prediction based algorithm by replacing the original rule-based bitrate adaptation logic by these implementations which were fully described by \cite{MPC}. We implemented our algorithm and made it accessible by AbrController class. 

We compare our algorithm (FastScan) with the FastMPC, Pensieve, Bola, and the other existing buffer and prediction based algorithms which were described by \cite{MPC}. We set the maximum buffer size to 1 minute for all considered algorithms $B_m=1 minute$.We briefly describe the comparable approaches here, and for more information about these algorithms, see \cite{MPC}.

\BULLET FastMPC: The FastMPC implementation has a static table
that is used to index control decisions. It uses
the harmonic mean to predict the  available bandwidth. We use default setting proposed by \cite{MPC}. The prediction window is of size $5$ chunks (i.e. look-ahead horizon $h = 5$) with throughput predictions using harmonic mean of past 5
chunks. We use 100 bins for throughput prediction and
100 bins for buffer level.

{\BULLET Pensieve: Pensieve is a Reinforcement Learning (RL) based rate adaptation scheme in which an RL agent is trained offline using real bandwidth traces to make quality decision per chunk. Once the RL is trained, it is used to make quality decisions for streaming video chunks. Pensieve uses the QoE metric that is proposed in \cite{MPC} as its reward function. we run Pensieve using the trained model provided by \cite{mao2017neural}.

\BULLET  FESTIVE: \cite{jiang2012improving}: It calculates efficiency score depending on throughput predictions using harmonic mean of the past 5 chunks, as well as a stability score as a function
of the bitrate switches in the past 5 chunks. The bitrate
is chosen to minimize stability score$+\alpha$ $*$ efficiency score. Where $\alpha=12$. 

\BULLET  BBA: Proposed by~\cite{BBA}, it adjusts the streaming quality based on the playback buffer occupancy.
Specifically, it is configured with lower and upper buffer thresholds (reservoir and cushion). If the buffer occupancy is lower (higher) than the lower (higher) threshold, chunks are fetched at the lowest (highest) quality; if the buffer occupancy lies in between, the buffer-rate relationship is linear. we set reservoir $r = 10s$ and cushion $c = 30s$.

\BULLET   RB: The bitrate of the next chunk is picked as the maximum available bitrate
which is less than throughput prediction using harmonic mean of past 5 chunks.

\BULLET   Bola: The original dash.js implementation that adopts a rule-based
bitrate decision logic as shown in Fig.~\ref{fig : dashjsSys}.

%
%have been post-processed in \cite{yin2015control} to give 1000 traces, each of 6-minute length which will be used in this paper for evaluations. %We, however, scaled 

\begin{figure}
\centering
\vspace{-0.25in}
\includegraphics[trim=0in 0in 0in 0in, clip,width=0.4\textwidth]{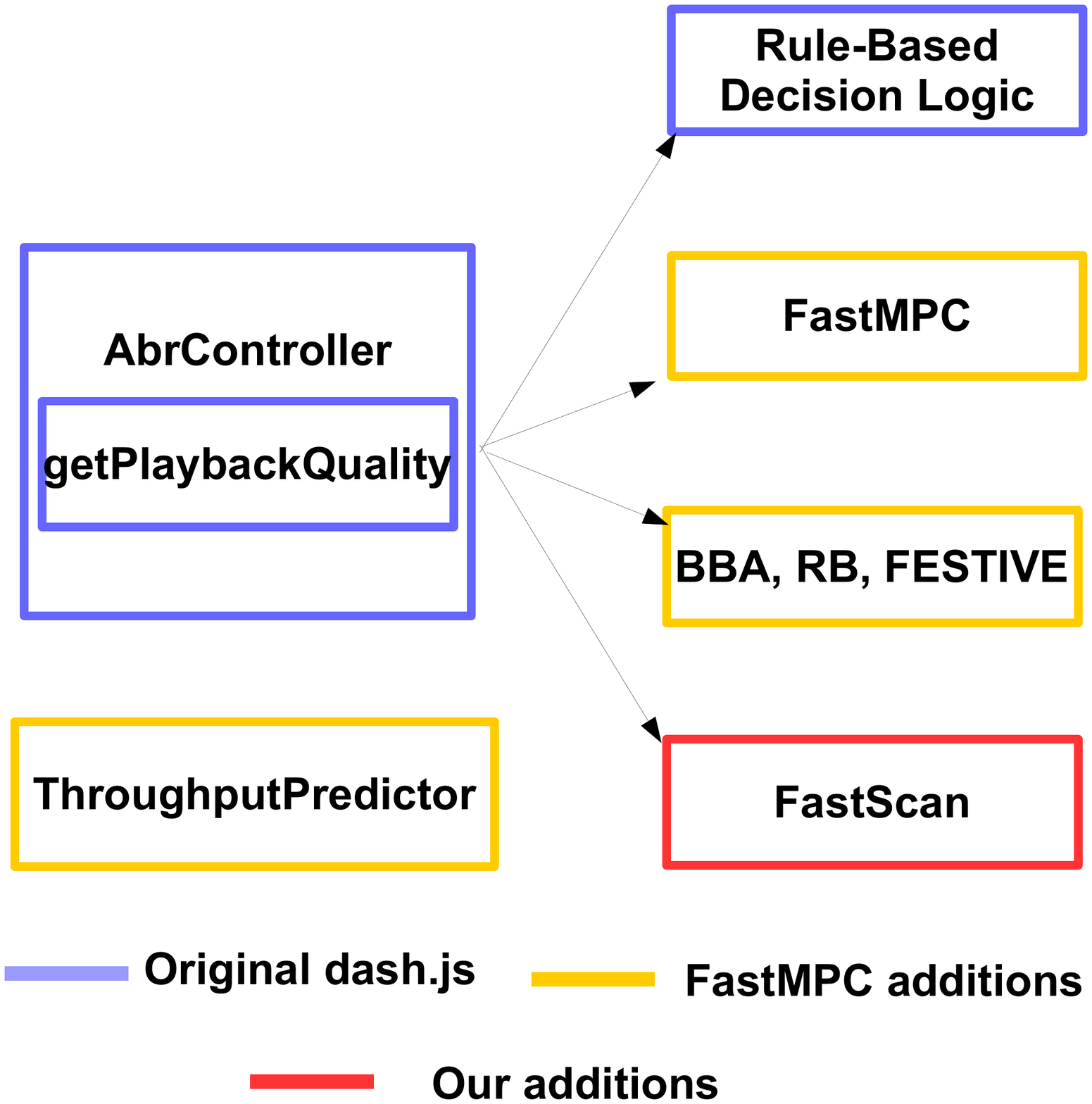}	
\vspace{-0.7in}
 \caption{dash.js code structure, FastMPC modifications \cite{MPC}, and our modifications}
 \label{fig : dashjsSys}
 \vspace{-.2in}
 \end{figure}

To make the comparison fair, we have the following settings for our algorithm: the prediction window is 5 chunks ($W=5$), we use the time corresponding to the download of the last 5 chunks to predict the future available bandwidth ($\eta=5$). Moreover, we use the nominal chunk size of all levels (not the exact chunk size) when we run the forward-backward algorithms. Therefore, we consider the performance when we have knowledge about the mean chunk sizes. We believe knowing the exact size will further improve the performance of our algorithm. Finally, we set the lower buffer threshold to $5s$ which is a little higher than playback time of 1 chunk. In other words, if the buffer does not contain a chunk, and the FastScan decision about the next chunk size is higher than the $1$-st quality level, we reduce the quality decision that is made by FastScan algorithm by 1 quality level.

{\bf Video parameters:} We use the Envivio video from DASH-
264 JavaScript reference client test page \cite{dasha}, which is 260s
long, consisting of 65 4s chunks ($L=4s$). The video is encoded by
H.264/MPEG-4 AVC codec into 5 different quality levels, and the nominal bitrate of the levels are as follows: $R = \{0.338$Mbps, $0.583$Mbps, $0.959$Mbps, $1.898$Mbps, $2.806$Mbps$\}$. The actual encoding rates of the chunks at any of the 5 different levels are different from the nominal rates since the video is encoded in variable bit rate (VBR). The min, mean, and max chunk size of every quality level in Mega Bytes (MB) are shown in Table~\ref{tab : vbr_rates} (chunk size of level $i$ is equal to $L*R(i)$). The table clearly shows the variability of the chunk sizes of every level. %Finally, we choose the maximum buffer size to be 1 minute ($B_{max} = 1$ minute). %\textcolor{red}{Vaneet: I have a question here!}

\begin{table}[htb]
  \centering
  \caption{AVC encoding chunk sizes in MB of the Envivio video used in our evaluation}
  \begin{tabular}{|c|ccccc|} \hline
    Quality level & level-0 & level-1 & level-2 & level-3 &level-4 \\ \hline
    Min & 0.0433 & 0.0786 & 0.1265 & 0.2213 & 0.3205 \\ \hline
    Mean & 0.1693 & 0.2916 & 0.4795 & 0.949 & 1.403 \\ \hline
    Max & 0.2342 & 0.3855 & 0.6217 & 1.286 & 1.918 \\ \hline
  \end{tabular}
  \label{tab : vbr_rates}
 % \vspace{-.1in}
\end{table}

{\bf Bandwidth Traces: } For available bandwidth traces, we used traces from three sets of public dataset. The first dataset consists of continuous 1-second measurement of throughput of a moving device in Telenor's mobile network in Norway \cite{riiser2013commute}. The dataset is post-processed by \cite{MPC} to make short traces that matches the video length. We avoid the traces in which the quality decision is trivial (\ie the very low traces in which the decision is to fetch all chunks at the lowest quality level, and the very high average traces, in which the decision is to fetch chunks at highest quality level). The second dataset is the  FCC dataset \cite{fccData}, which consists of more than 1 million sets of throughput measurements. The last dataset consists of 40 LTE traces collected in Belguim \cite{vanderHooft2016}. Since LTE traces have high bandwidth values, we scale every trace by a factor of $1/5$ to create more challenging scenarios. %\textcolor{red}{Vaneet: Can you read this paragraph and comment if any}
\if0
\begin{figure}
\input{bwStatV2.tex}
 \vspace{-.1in}
 \caption{Statistics of the two available bandwidth traces: (a) mean, and (b) standard deviation of each trace's available bandwidth.}
 \label{fig : bwStatV2}
 \vspace{-.1in}
 \end{figure}
\fi

\input{evaluation}

%% file: evaluation.tex
\subsection{Evaluation}
In this section, we compare our approach against several prior approaches.
Our basic experiment setup consists of two computers (Ubuntu
12.04 LTS) emulating
a video server and a client. The video client is a GoogleChrome
web browser with V8 JavaScript
engine while the video server is a simple HTTP server based
on node.js (version 0.10.32). We use the dummynet \cite{dummynet} tool
to throttle the throughput of the link between two computers
according to the throughput traces employed.

\begin{figure}
\centering
%\begin{figure}
\vspace{-.05in}
\includegraphics[trim=0.2in 0in 0.5in 0in, clip,  width=0.48\textwidth]{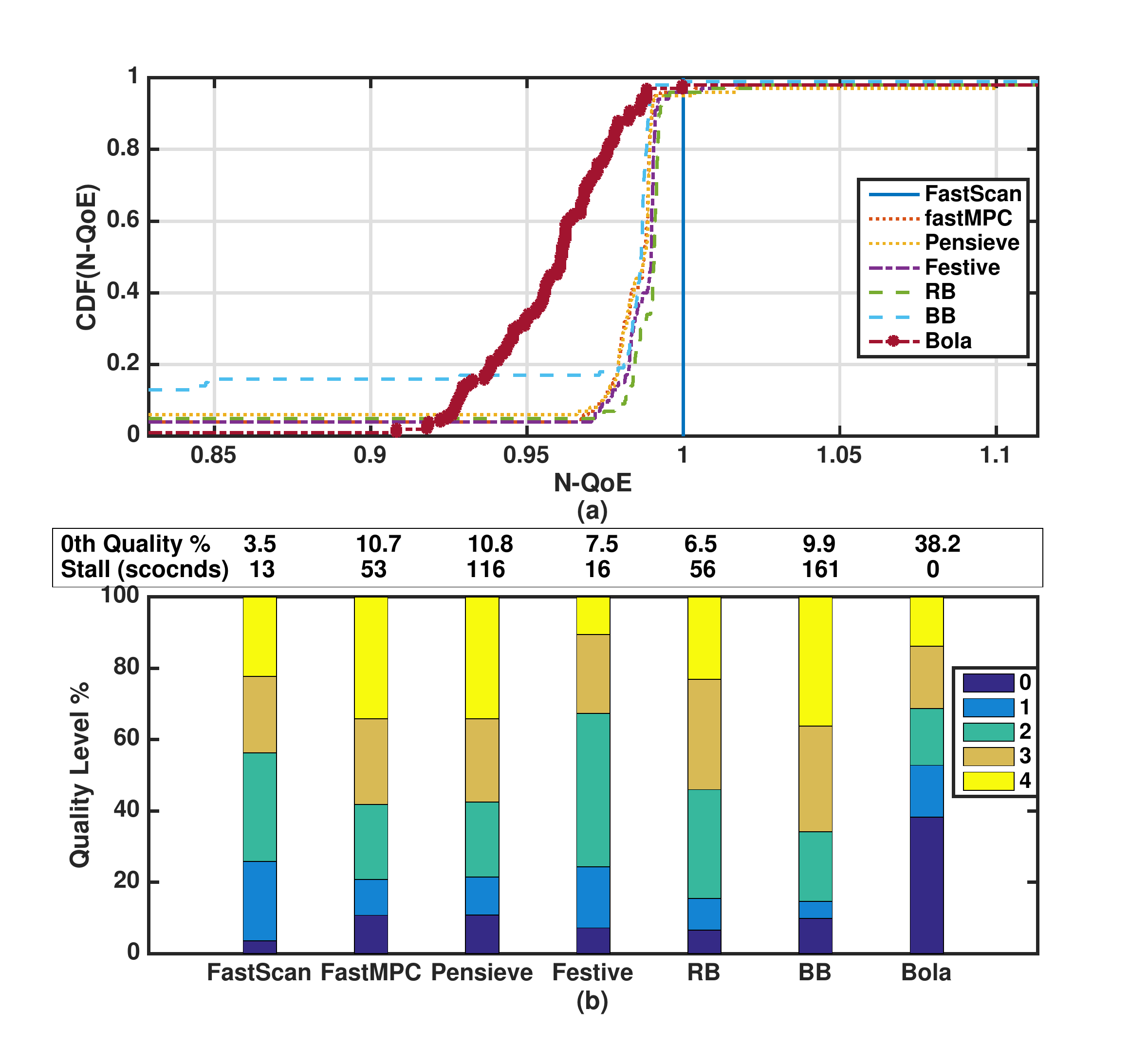}	
\vspace{-.1in}
\caption{ Comparison with other approaches: Norway Dataset}
 \label{fig : BCompeq-hsdpa}
\end{figure}

\begin{figure}
\centering
%\begin{figure}
\vspace{-.05in}
\includegraphics[trim=0.2in 0in 0.5in 0in, clip,  width=0.48\textwidth]{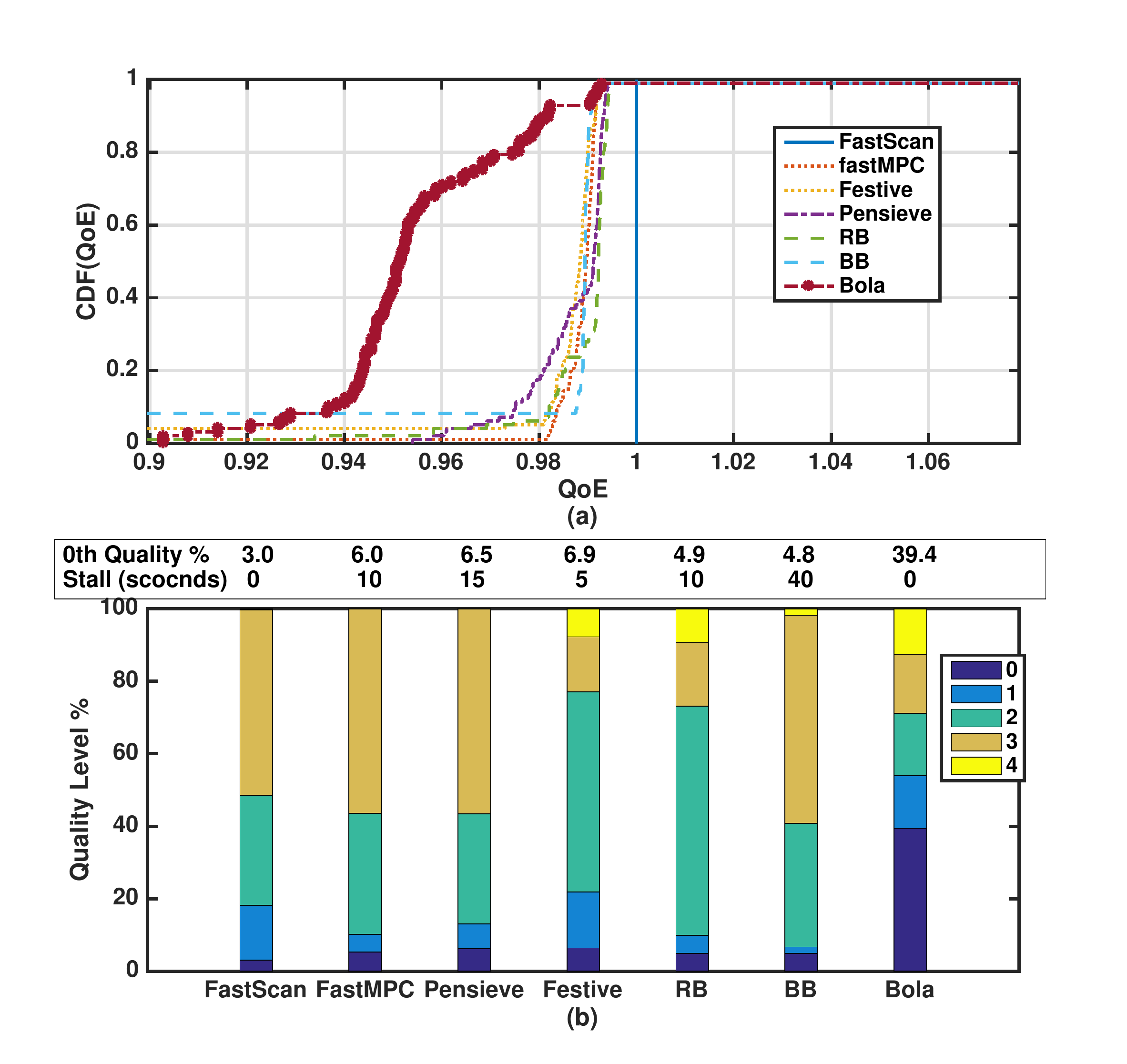}	
\vspace{-.1in}
\caption{ Comparison with other approaches: FCC dataset}
 \label{fig : BCompeq-fcc}
\end{figure}

\begin{figure}
\centering
%\begin{figure}
\vspace{-.05in}
\includegraphics[trim=0.2in 0in 0.5in 0in, clip,  width=0.48\textwidth]{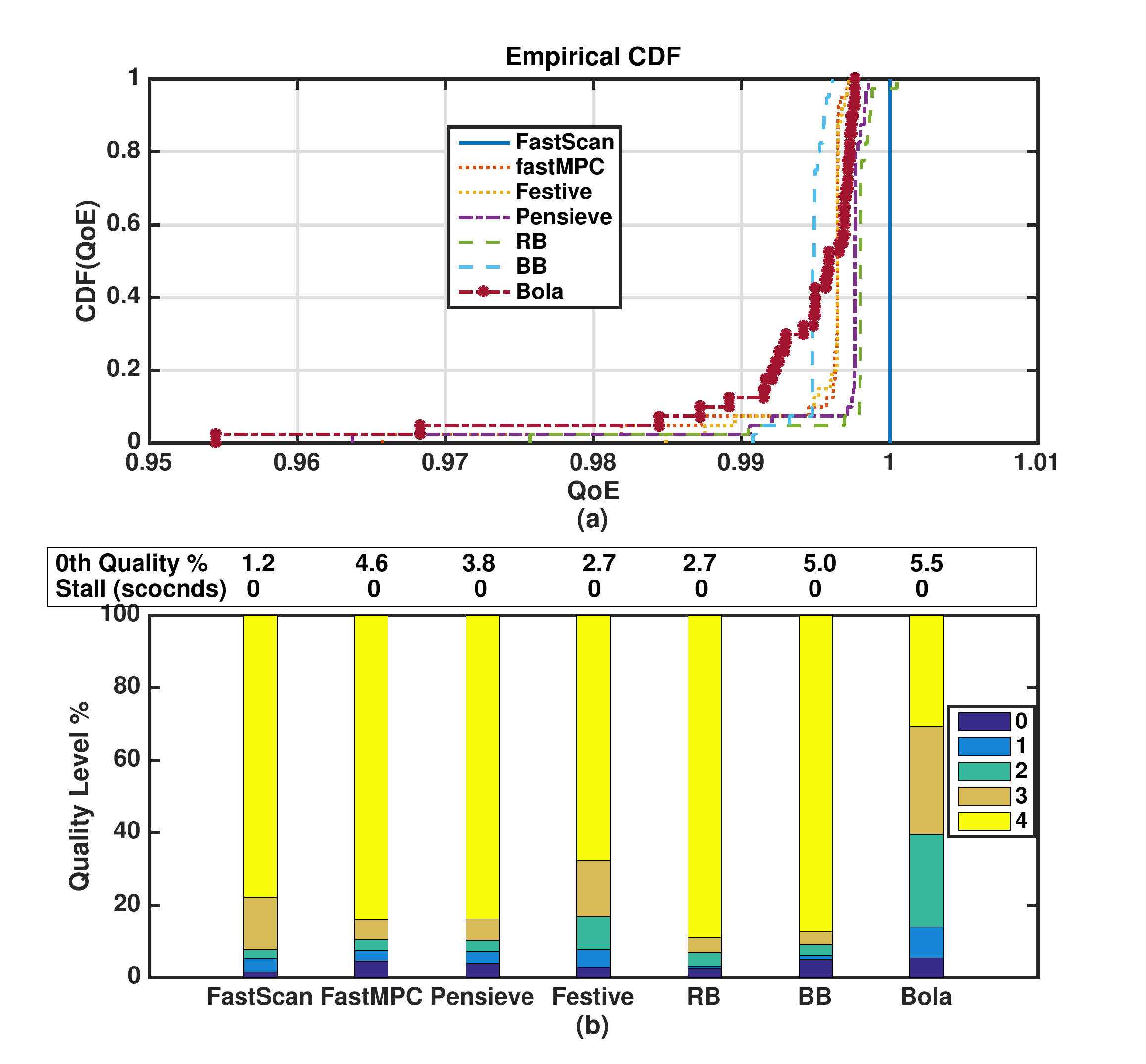}	
\vspace{-.1in}
\caption{ Comparison with other approaches: LTE dataset}
 \label{fig : BCompeq-LTE}
\end{figure}

\if0
\begin{figure}
\centering
%\begin{figure}
\vspace{-.05in}
\includegraphics[trim=0.2in 0in 0.7in 0in, clip,  width=0.48\textwidth]{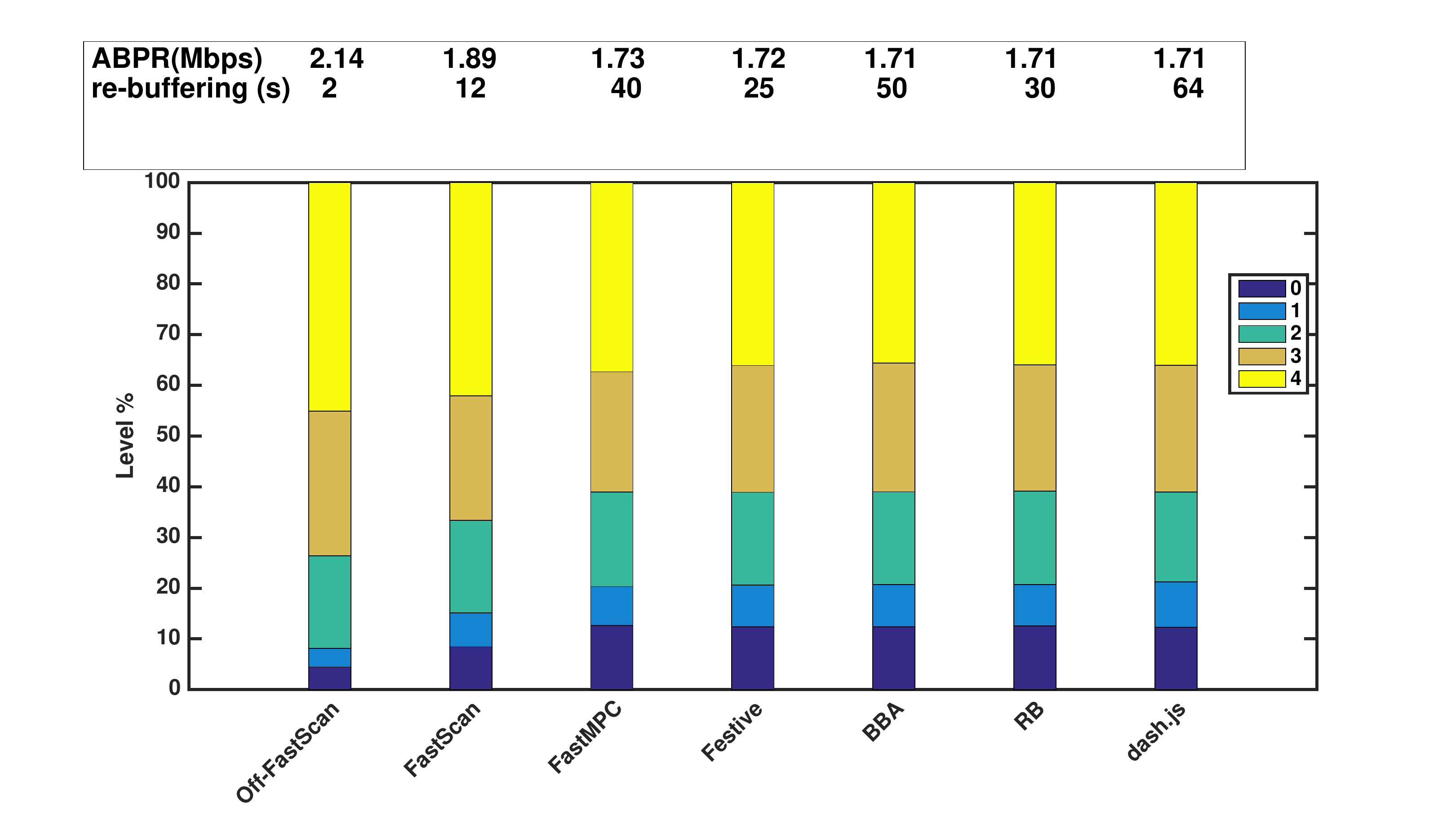}	
\vspace{-.1in}
\caption{ Comparison with other approaches: Video Quality distribution}
 \label{fig : BCompeq}
\end{figure}
%\vspace{-0.4in}
\begin{figure}
\centering
%\begin{figure}
%\vspace{-.1in}
%\input{avgCDF_final.tex}
\includegraphics[trim=0.5in 0.5in 0.5in 0.3in, clip,  width=0.48\textwidth]{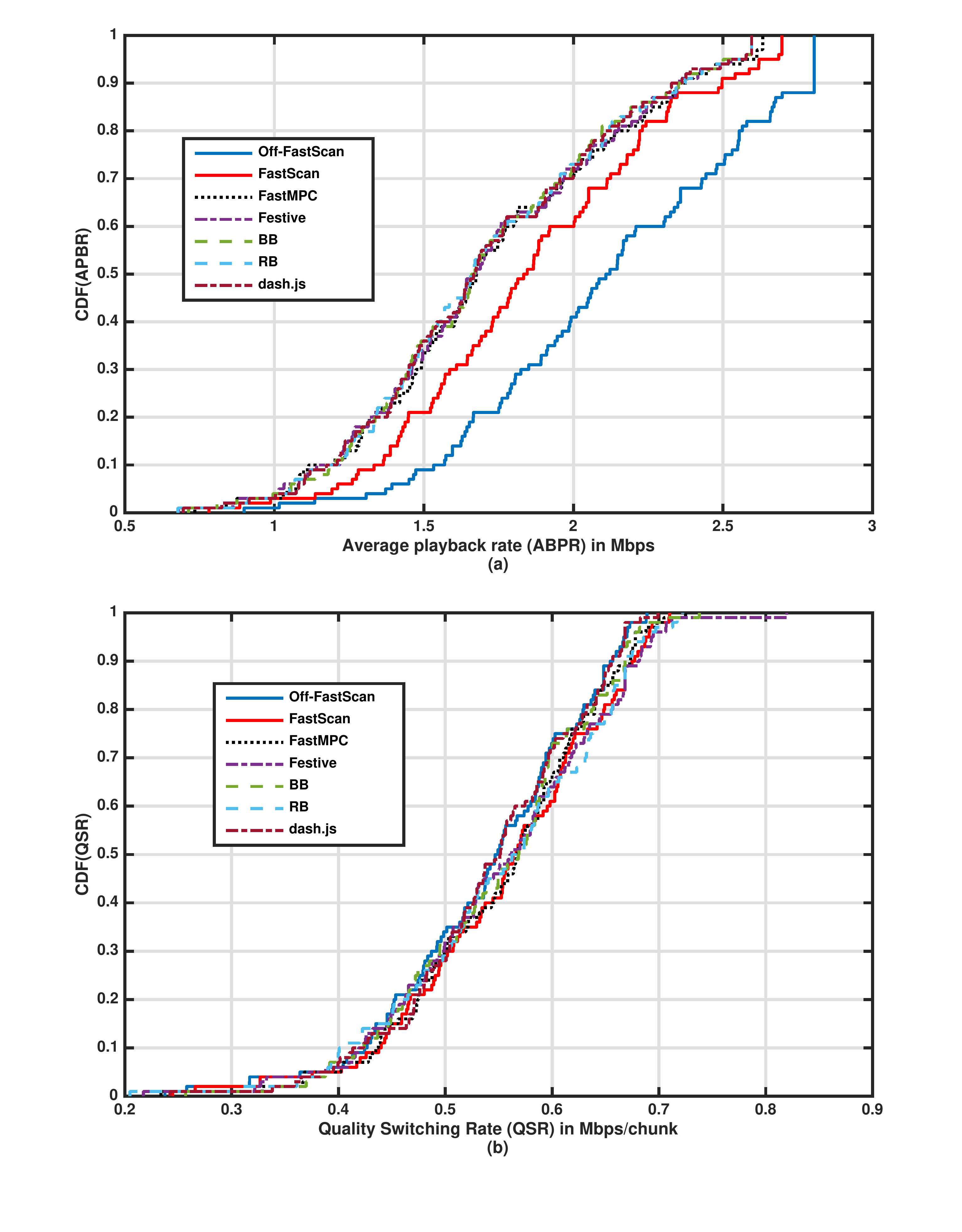}	
\vspace{-.2in}
\caption{ Comparison with other approaches: (a) CDF of Average Playback Rate (APBR), (b) CDF of Quality Switching Rate (QSR).}
\vspace{-.2in}
 \label{fig : BCompeq2}
\end{figure}
\fi

%\begin{thisnote}
	The results are shown in Figures ~\ref{fig : BCompeq-hsdpa} -- \ref{fig : BCompeq-LTE}. Fig.~\ref{fig : BCompeq-hsdpa}-a shows the CDF of the normalized QoE. The QoE metric is our objective function when $\beta=0.1$ and $\lambda=10$. The normalization is with respect to FastScan's QoE. We clearly see that in the HSDPA dataset (Norway dataset), our algorithm acheives the highest QoE for about $99\%$ of the bandwidth traces, and the gain is significant in some of the traces. To get more insights we plot in Fig.~\ref{fig : BCompeq-hsdpa}-b the probability mass function of the number of chunks fetched at the different quality levels (e.g ``0"= the lowest quality level, and ``4"= the highest quality level). 
	
	The percentage of of the chunks that are downloaded at the poorest quality ($0$-th quality level) and the total re-buffering (stall) duration for each algorithm among all traces is displayed on the top of Fig.~\ref{fig : BCompeq-hsdpa}-b. We first see that FastScan outperforms most of the baseline algorithms in terms of avoiding stalls (re-buffering durations). For instance, over 100 available bandwidth traces, FastScan runs into 13 seconds of buffering time (stalls). On the other hand,  FastMPC, the algorithm that achieves the closest average playback to FastScan runs into 53 seconds of the stall time which is about 4 times higher than the FastScan's stall duration. We clearly see that FastScan maintains a very low stall duration and pushes more chunks to higher quality levels. FastScan fetches only $3.4\%$ of the chunks at the poorest ($0$-th) quality level which is about half of the Festive's percentage and one third  of FastMPC and Pensieve. FastMPC and Pensieve fetch more chunks at the highest quality level than FastScan, but that comes at the cost of fetching more chunks at the poorest quality level and running into more stalls. However, since the considered QoE metric is a concave function with respect to the playback rate, pushing more chunks from $0$-th to the $1$st quality level achieves higher QoE than fetching more chunks at the highest quality level at the cost of dropping the quality of some chunks to the $0$-th quality level. Therefore, FastScan achieves higher QoE since it processes quality levels in order and maximizes the number of candidate chunks to the $n$-th quality level before considering the $(n+1)$ quality level. It worth mentioning that Bola achieves the minimum stall duration (0 seconds), but that comes at the cost of achieving significantly lower QoE in  most of the bandwidth traces and the highest percentage of chunks that are fetched at $0-th$ quality level ($38.2\%$). Finally, FastScan incorporates available bandwidth prediction and the deadline of the chunks into its decisions, prioritizes the later chunks, and re-considers the decisions periodically (after the download of every chunk). These properties help FastScan be adaptive to different available bandwidth regimes and variations in the available bandwidth profiles. 
	
	Fig.~\ref{fig : BCompeq-fcc} compares the results of  FastScan with the other algorithms in FCC dataset,  and Fig.~\ref{fig : BCompeq-LTE} compares the results of  FastScan with the other algorithms in LTE datset. In both datasets, we see qualitatively similar results to what we have described in HSDPA dataset. In the two set of traces, our algorithm achieves  highest QoE in more than $99\%$ of the traces. Moreover, our algorithm is able to manage the trade-off between pushing more chunks to higher quality levels, and avoid running into stalls. It maintains a stall duration that is as low as the one achieved by the  most conservative algorithm and high average playback as high as the one that is achieved by the most optimistic algorithm.

	In conclusion, The proposed algorithm (FastScan) significantly outperforms the considered baselines in terms of avoiding stalls and pushing more chunks to their highest quality levels. Incorporating the predicted bandwidth and the deadline of the chunks into its decisions, prioritizing the later chunks, and re-considering the decisions after the download of every chunk make FastScan adaptive to different available bandwidth regimes and variations in the available bandwidth profiles. 
	
%\end{thisnote}

%% file: conclusion.tex
 \section{Conclusion}
 
 This work considers the problem of adaptive bit-rate video streaming, which optimizes a novel QoE metric that models a combination of the three objectives of minimizing the stall/skip duration of the video, maximizing a concave function of the playback quality averaged over the chunks, and minimizing the number of quality switches. A low-complexity algorithm is proposed to solve the optimization problem. Due to the structure of the proposed QoE, the proposed algorithm can be shown to be optimal under some assumptions. Extensive evaluations with real videos and  available bandwidth traces of a public dataset reveal the practicality and the robustness of the proposed scheme and demonstrate its significant performance improvement as compared to the state-of-the-art ABR streaming algorithms.
 
%In this work, a non-convex optimization problem that optimizes a novel QoE metric that models a combination of the three objectives of minimizing the stall/skip duration of the video, maximizing a concave function of the playback quality averaged over the chunks, and minimizing the number of quality switches is proposed. A low-complexity algorithm is proposed to solve the optimization problem, and it is shown to achieve the optimal solution under some assumptions. The algorithm considers the predicted available bandwidth, the deadline of chunks, and the buffer occupancy to make a  quality decision for each chunk. Extensive evaluations with real videos and  available bandwidth traces of a public dataset reveal the practicality and the robustness of the proposed scheme and demonstrate its significant performance improvement as compared to the state-of-the-art ABR streaming algorithms.

%Extension of the approach to utilize multiple paths (e.g. WiFi and LTE) is an important problem. The authors have developed some preliminary ideas on this topic in \cite{spcom}. 